\documentclass[a4paper,fleqn]{cas-sc}
\usepackage{setspace}
\usepackage[switch]{lineno}
\usepackage[numbers]{natbib}
\usepackage{nccmath}
\usepackage{lineno}
\usepackage{float}
\usepackage{amsmath}
\usepackage{parskip}
\usepackage{caption} 
\usepackage{upgreek} 
\usepackage[section]{placeins}

\begin{document}

\let\WriteBookmarks\relax
\def\floatpagepagefraction{1}
\def\textpagefraction{.001}
\shorttitle{}
\shortauthors{D. Palo, W. Molzon et~al.}

\title [mode = title]{Neural Network Applications to Improve Drift Chamber Track Position Measurements}                

\author[1]{D. Palo}[orcid=0000-0001-9256-5348]
\cormark[1]
\ead{dpalo@uci.edu}
\author[1]{W. Molzon}[orcid=0000-0001-9232-6064]

\address[1]{University of California, Irvine, CA 92697}


\begin{abstract}
This paper describes applications of two neural networks to improve drift chamber  position measurements. One network calculates a data-driven estimate of the drift cell time-to-distance relationship that is conventionally estimated by a numerical calculation based on the anode and cathode wire geometry, wire potentials, and gas properties.  The second network additionally uses the full digital waveform of the signal in the drift chamber, hence accessing information on the full ensemble of ionization clusters. This network uses more information than the conventional position estimate that relies exclusively on the arrival time of the first drift electron. In principle, this technique improves resolution even when multiple ionization clusters cannot be separated, in contrast with a cluster-counting technique. The performance of both networks when applied to MEG II drift chamber data is reported and compared to that of a conventional approach.  
\end{abstract}

\begin{keywords}
\vspace{-6px}
Drift Chamber \vspace{-6px} \sep 
Machine Learning \vspace{-6px} \sep
Convolutional Neural Network \vspace{-6px} \sep 
Cluster Counting\vspace{-6px} 
\end{keywords}

\maketitle

\doublespacing

\section{Introduction}
The primary position measurement of an ionizing particle traversing a drift chamber is the distance of closest approach (DOCA) of the particle to the cell's sense wire. Optimizing the accuracy and precision of the measurement is critical to a variety of modern day particle physics experiments ranging from high energy lepton collider experiments to rare decay searches at low momentum. In many applications, the positron measurement resolution in a magnetic spectrometer has significant contributions from both the position measurement and scattering in the drift chamber material. The position resolution has contributions from ionization statistics, longitudinal diffusion, electronic noise, contamination from signals due to an out-of-time ionizing particle, etc. This paper focuses on improving the position measurement by applying neural network techniques. In Section \ref{Conventional}, the conventional approach to estimating the DOCA in a drift chamber is discussed. In Section \ref{Improve}, we discuss limitations to the conventional DOCA estimation and how using a neural network might improve it. 
In Section \ref{MEGII}, we give a brief overview of the MEG II experiment and an explanation of how the neural network application could improve its DOCA estimation. We then present the neural network methods in Section \ref{NN}. In Section \ref{Results}, we discuss the improvements in kinematic measurements resulting from applying the neural network approach to  MEG II drift chamber analysis. Finally, in Section \ref{Discuss}, we discuss potential improvements to the technique.


\subsection{Conventional DOCA Estimation}
\label{Conventional}
 Drift chamber analysis is generally initiated by searching for drift cells intersected by potential particle trajectories (tracks). The digitized time dependent signal (waveform) of such a particle passage is referred to as a hit. The analysis then reconstructs particle position information including the DOCA estimate. This information is fed to a track finding algorithm to cluster hits into particle tracks. Finally, a Kalman filtering method\cite{Kalman1}\cite{Kalman2} is commonly used to fit the tracks and provide a kinematic estimate.

The DOCA calculation starts by estimating the arrival time of each hit's first ionization cluster (clusters contain typically 1-2 ionization electrons) in the waveform. This is converted to a drift time by determining the particle time at the cell (referred to as the $T_{0}$) from an independent measurement. This $T_{0}$ is either from a signal in a fast scintillation counter or from a fit to multiple drift chamber hits on the particle trajectory. The drift time is then converted to the DOCA using the position dependent drift velocity that is a calculated or measured value, which depends on the local electric and magnetic fields in the cell and on the gas mixture used. The drift velocity is either separately measured or calculated using a code such as Garfield++\cite{Garfield}. Garfield++ estimates the time-distance relation (TXY function) between the ionization position in the cell and the drift time by simulating the pattern of ionization sites in the gas mixture, the cell's electric field, and accounting for the external magnetic field. In cases (such as the MEG II drift chamber\cite{CDCHDetector}) where the drift cell is not circular, this conversion depends on the track angle in the cell and actually provides a position of closest approach. In either case, two such points are provided corresponding to the particle being on one or the other side of the wire. This ambiguity (conventionally called the left right ambiguity) is resolved in the fitting process. 

 In practice, the DOCA estimate is derived iteratively. Before any tracking, the drift time is converted to a DOCA averaging over all track angles. After tracking, the DOCA is reevaluated with the estimated track angle. To estimate the $T_{0}$ before tracking, a nominal propagation delay from the drift cell to a fast scintillating counter is used.  After tracking, the propagation delay and thus the $T_{0}$ at each drift cell is known precisely.

In practice, several complications affect this process. Determining the arrival time of the first cluster requires finding the time of a possibly low amplitude signal in the presence of electronics noise. Further, the amplification of individual drift electrons has a distribution (typically described by Polya functions\cite{Polya}) with significant probability of having values well below the average. Depending on the values of the Polya function parameters, the function varies from nearly exponential to Gaussian. The distribution is not easily calculated or measured for complicated gases.  This procedure is further complicated in high rate experiments in which the hit waveform may contain contributions from more than one track.




\subsection{Improving the Conventional DOCA Estimate}
\label{Improve}
Here, we list the issues with the conventional DOCA estimate that may contribute to systematic errors in the most probable value and dispersion of the DOCA estimate, and how using a neural network avoids or minimizes these effects.  
  
\begin{figure}[H]
{\centering
\includegraphics[width=8cm]{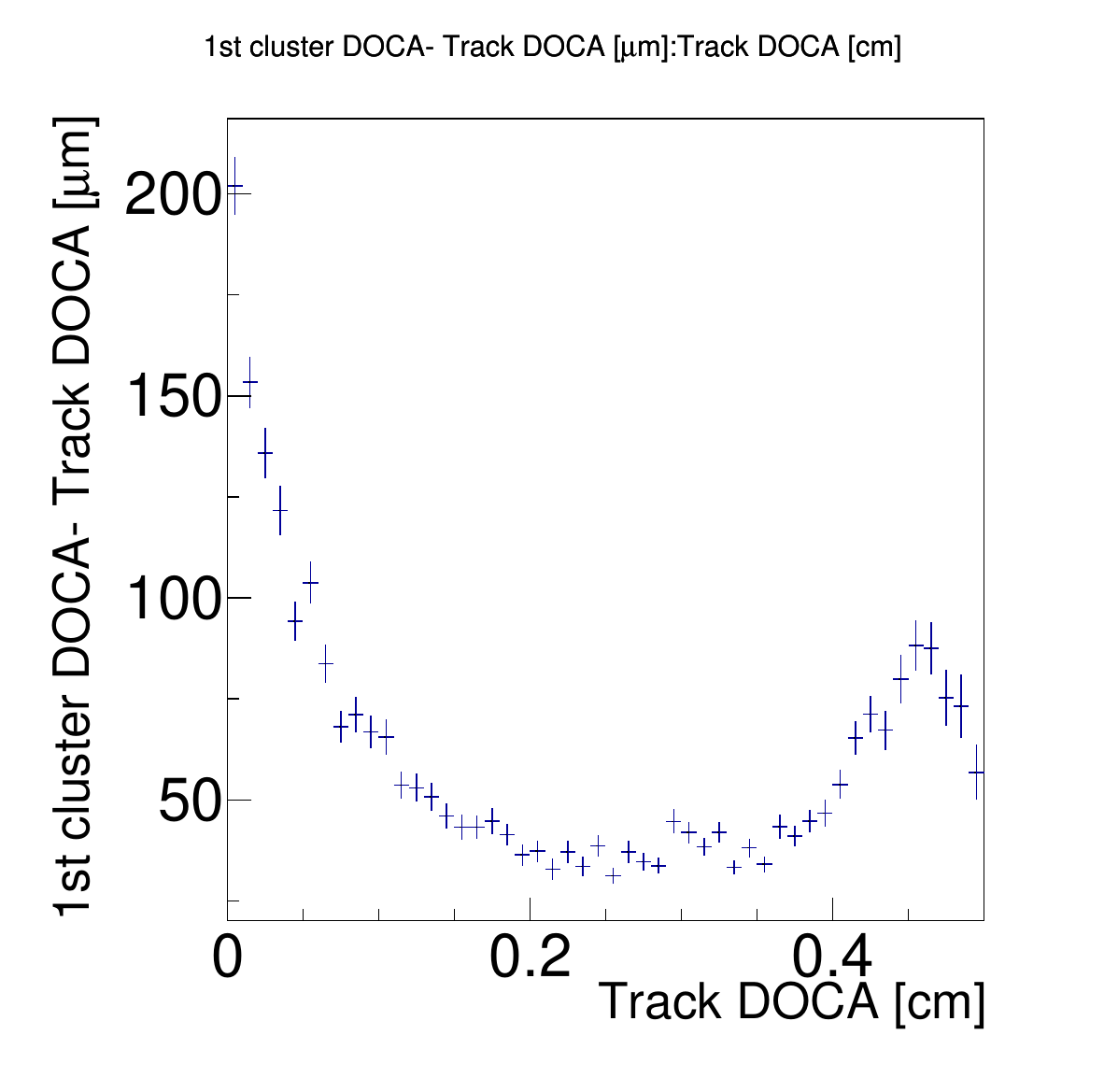}
\caption{The mean value of the difference between the first ionization cluster position and the distance of closest approach (DOCA) of the track to the sense wire (DOCA bias) as a function of the track DOCA, from a Geant4 based simulation of the MEG II cylindrical drift chamber. It is summed over all wires to show the general trend of the DOCA bias. This bias shows the ionization statistics effect in the gas mixture. 
\label{DocaBiasMC}
}}\end{figure}

First, the conventional approach systematically overestimates the DOCA in any drift chamber with a low cluster density.  This is illustrated in Figure \ref{DocaBiasMC}, which shows the mean DOCA bias (i.e. DOCA of the 1st ionization cluster  - the track DOCA) as a function of the track DOCA in the MEG II Geant4-based simulation\cite{Geant}. At small DOCA (<2 mm), the first ionization site is farther from the wire than the track DOCA by 50-200 $\mu$m. By training on data, the neural network approach can learn and thus remove this bias as a function of drift time, track angle, cell size, etc. 

Next, in many analyses, including that of MEG II, the DOCA estimate relies on Garfield++ to accurately estimate the ionization pattern/statistics and the drift velocity in the gas mixture in use. This comes with uncertainty that is eliminated by a data-driven TXY function. 

Garfield++ is inherently two dimensional; if the cell size or relative anode cathode geometry is changing along the wire axis, the algorithm necessitates an interpolation of the TXY function from calculations at fixed positions along the wire axis. Additional uncertainty in the TXY function arises from assembly variations in wire positions and the effects of bowing of wires from electrostatic and gravitational effects. These variations in the TXY function can in principle be corrected in the machine learning analysis given a sufficiently large training sample. 

Finally, the conventional DOCA estimate uses only the drift time of the first (primary) drift electron. An improved DOCA resolution can be achieved by identifying ionization clusters and their times (i.e. cluster counting)\cite{Cluster Counting 1}\cite{Cluster Counting 2}. Cluster counting algorithms map the cluster time distribution to an improved final DOCA estimate relying on a correct determination of the number and pattern of clusters in a hit. In drift chambers with a low signal/noise ratio, it is difficult to distinguish overlapping clusters and differentiate noise from low amplitude clusters. Since all available ionization cluster information is stored in the digital waveform, a neural network may use the available information without explicitly counting clusters to determine the best DOCA estimate.

\section{Methods}
\subsection{Example Implementation: MEG II Experiment}
\label{MEGII}

As an example, the neural network algorithms are applied to data from the MEG II drift chamber. The MEG II experiment\cite{MEG II} is an active search for the charged lepton flavor violating\cite{CLFV} decay of an anti-muon ($\upmu^{+}$) to a positron ($e^{+}$) and a photon ($\gamma$): $\upmu^{+} \rightarrow e^{+} \gamma$. Results from the first MEG II physics run operation and performance are described here\cite{MEGIIDetector}. The experiment proceeds by stopping $\upmu^{+}$ at $\sim 4\cdot 10^{7}$ Hz in a thin target and then measuring the kinematics of the muon decay products. The positrons are measured in a cylindrical drift chamber (CDCH) and a set of pixelated scintillation counters (SPX). The COnstant Bending RAdius (COBRA) superconducting solenoid provides the field for the magnetic spectrometer. 
The field is graded to reduce the number of turns in the solenoid by particles emitted nearly normal to the solenoid axis, which would otherwise cause excessive rates in the CDCH.
The neural network applications are applied to the CDCH data. A complete description of the CDCH analysis details and performance are described here\cite{CDCHDetector}. 


 The experiment's signal and background have the same observable particles; hence unambiguous evidence of the signal depends critically on measuring the kinematics of the decay products with high precision. For the positron, the kinematic measurements include the momentum, emission angle, vertex position, and time at the target. 
 The neural network approaches shown in this paper improve the single hit resolution and therefore the positron kinematic measurements and hence the experiment's background rejection and sensitivity. 



The CDCH consists of 1728 drift cells filled with $He$:$C_{4}H_{10}$:$O_{2}$:$C_{3}H_{8}O$ (88.2:9.8:0.5:1.5). Details on the design and construction of the chamber are found here\cite{MEG II}. The sense wires are oriented in alternating stereo angles of $\pm \sim 7^{o}$ degrees with respect to the chamber axis to enable reconstructing a track's axial coordinate. 

\begin{figure*}[b]
{\centering
\includegraphics[width=12cm]{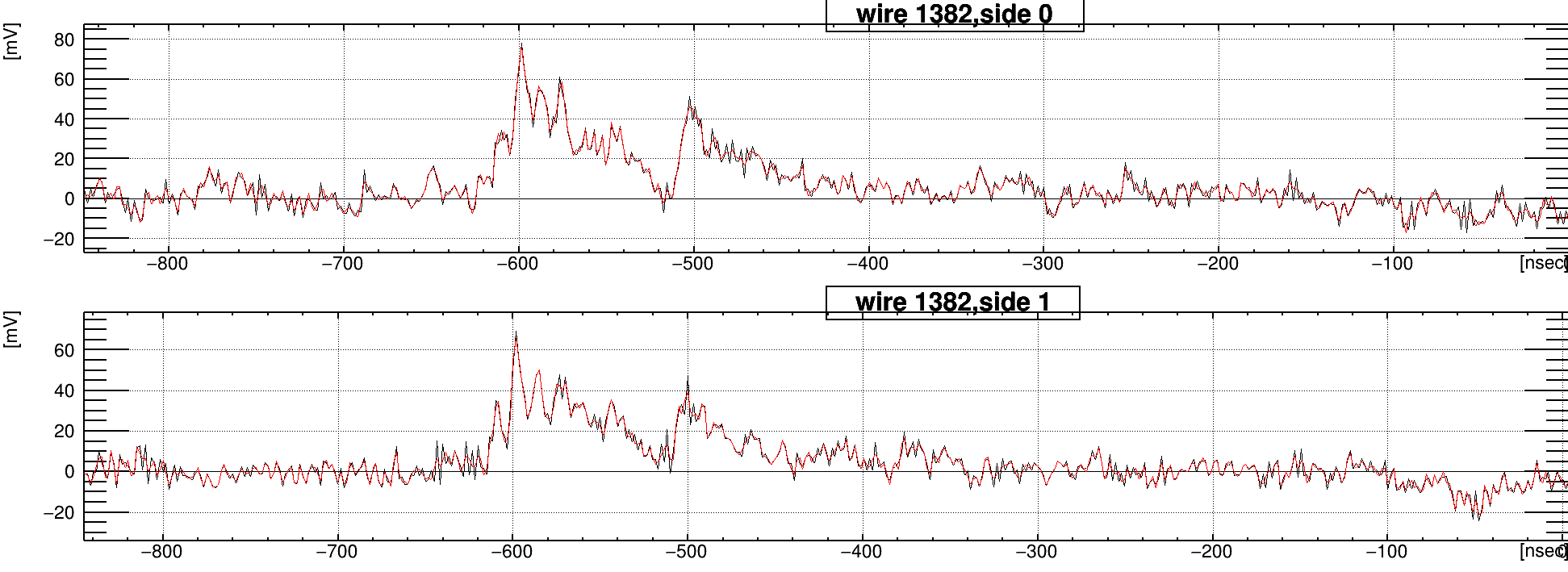}
\caption{A typical CDCH digitized waveform. The black curve already has low frequency noise suppression applied. The red curve has both high and low frequency noise suppression applied.
\label{waveform}
}}\end{figure*}

The drift cells are approximately square with a mean full width of $\sim$ 8 mm, varying layer-by-layer and by position along the wire. Signals on the wire are digitized at $\sim 0.6$ GHz by electronics \cite{wavedream} on both ends of the wire, providing a preliminary estimate of the hit position along the wire by the relative charge and time at the two ends. Due to the stereo geometry, the position of the cathode wires with respect to the anode wires vary for different anode wires and as a function of the position along the anode wires. The maximum drift time is about 400 nanoseconds at the edge of the drift cell, varying with cell size. The isochrones (loci of positions with equal drift times) are not circular due to the approximately square distribution of field wires and the magnetic field. Hence, the TXY function depends on the track angle. Based on Garfield++ simulations\cite{Garfield}, a signal positron creates $\sim 15$ ionization clusters/cell, each with an average of $\sim 1.3$ electrons. A CDCH digital waveform of a typical intersected drift cell is shown in Figure \ref{waveform}.



Here, we list limitations of the conventional DOCA estimation specific to the MEG II experiment.

First, in the CDCH, the location of the cathode wires surrounding a given anode wire varies with position along the wire due to the stereo geometry in the drift chamber. In addition, the cell size is changing along the wire axis and layer-by-layer. Since Garfield++ is an inherently 2D program, the analysis relies on slices along the wire axis; the magnetic field and therefore the TXY function is then approximated along the wire. This explicit binning is not required by the neural network. 

Simulating any gas mixture comes with uncertainties; this is especially true when using a non-standard gas mixture like the 4-gas mixture used in MEG II. The standard MEG II Garfield++ simulation of the TXY function does not include oxygen or isopropanol; including them did not yield any clear improvement in the DOCA estimate. A description of the MEG II Garfield++ TXY is found here\cite{CDCHDetector}. Training a machine learning algorithm on data removes reliance on the simulation. Further, the isopropanol and oxygen lower the gas gain and therefore the signal/noise ratio in the waveform, worsening the precision with which the first ionization cluster time is measured and making estimating the number and time distribution of clusters (cluster counting) more difficult. 

\subsection{Neural Networks}
\label{NN}
In this section, we describe the two neural network approaches to DOCA estimation. First, a dense neural network (DNN) is trained to create a data-driven TXY function. That is, converting the waveform analysis hit arrival time into a DOCA estimate depending on other variables like cell size and track angle. Here, there is no explicit dependence on these other variables, but the network can implicitly learn how the DOCA estimation depends on these variables. Second, a convolutional neural network (CNN) also uses the full waveform to access information from all ionization clusters. The convolutional network layers are used to incorporate the temporal correlations between the waveform bins. 

\subsection{Training}

The networks are trained using MEG II data taken at low beam rate. We use the fitted track DOCA (unbiased by the hit itself) as an estimator of the true DOCA for the training. This technique results in a data-driven DOCA estimator. Both networks were trained in Keras\cite{Keras}.  The input variables ($X_{i}$) have been normalized using their mean ($\mu_{i}$) and standard deviation ($\sigma_{i}$): $X_{i} \rightarrow \frac{X_{i}-\mu_{i}}{\sigma_{i}}$. The track DOCA (Y) estimate has the same normalization:  $Y \rightarrow \frac{Y-\mu}{\sigma}$. This normalization is applied for optimal training (i.e. find the optimal network weights and thus the minimal loss function). The network was trained with a mean absolute error loss function ($\sum|Y-\hat{Y}|$) to avoid a large dependence on outliers (Y represents the fitted track DOCA and $\hat{Y}$ represents the network's DOCA estimate). 

 The use of low beam rate data ($1 \cdot 10^{6}$ Hz, $\sim 40$ times lower than standard beam intensity) effectively removes all out-of-time or "pileup" tracks that otherwise contaminate the signals. We found that using low intensity beam rate data for training improves the DOCA resolution even at standard beam intensity. This implies that the network has difficulty learning the TXY function at the same time as distinguishing between in and out of time signals (i.e. pileup discrimination). We come back to this point in Section \ref{Discuss}. 

Below, we list the input variables with some justification. All variables are included in both networks. Some of these variables are specific to the MEG II application. For example, if the cell size and relative anode cathode geometry was independent of the position along the wire axis, it would not be necessary to include this as an input variable. 

\paragraph{\textbf{Input Variables}}
\begin{itemize}
  \item Hit arrival time - This reconstructed hit time is the result of conventional waveform analysis and is the primary input to the TXY function.
  \item Layer number - The cell size decreases with layer number. There could be unique properties to specific layers (e.g. the first/final layers) so it's also included as a hot encoded variable (binary length 9 array).
  \item Wire number in the layer - The average track angle varies with the wire's global $\phi$. There may be trends in the anode cathode geometry as a function of wire number in a layer. 
  \item $T_{0}$ at the wire - This is required for any DOCA estimator. It is the SPX-based estimate and has an uncertainty, $\sigma_{T_{0}}$, of 
  35-90 ps with a full $T_{0}$ range of $\pm 10$ ns. 
  \item Track angle at the wire - Isochrones are not circular, thus the angle is required for the optimal estimate. 
  \item Kalman filter stereo reconstructed longitudinal wire coordinate - Cell size and thus the TXY function vary along the wire. 
  \item Waveform channel time and gain calibrations - These are useful for the convolutional network to normalize the wire-by-wire and channel-by-channel differences due to gain differences, cable lengths, etc.  
\item Binary flags to represent if a given wire end has been disconnected or is noisy - These ends are excluded from the analysis. 
 \item First stored waveform time bin - Each bin contains a stored voltage and time; this binned time is included to avoid the possibility of mapping the binned voltages to an incorrect drift time and thus an incorrect DOCA. 
\end{itemize}

\subsection{Dense Neural Network}
\begin{figure}[H]
{\centering
\includegraphics[width=8.7cm]{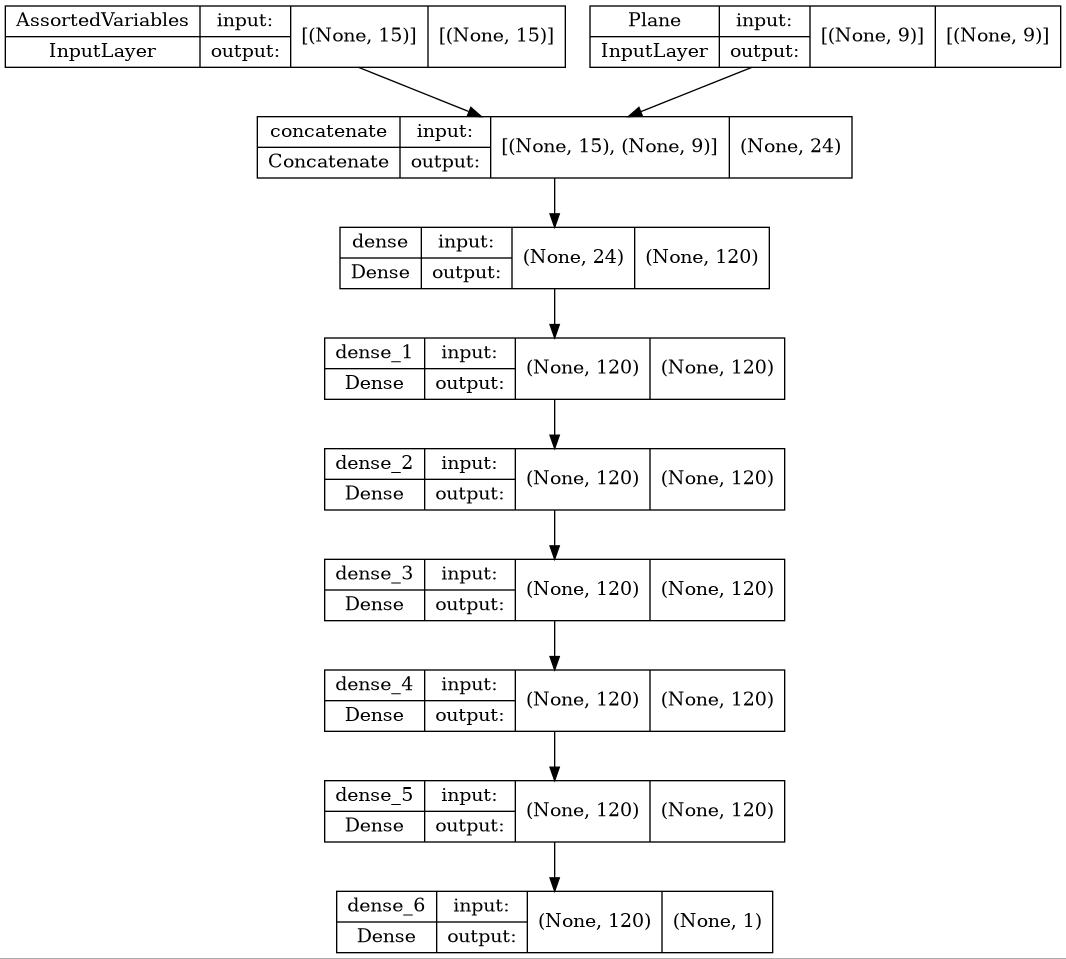}
\caption{The dense neural network architecture is shown. The neural network estimates the time-distance relationship using the hit arrival time and tracking information. 
\label{DNN}
}}\end{figure}

The dense neural network calculates a DOCA estimate given a hit arrival time; this is the TXY function. This network is shown in Figure \ref{DNN}. Since it is trained on data, it can learn the true TXY function as a function of the input variables. In addition, this network can learn the bias due to ionization statistics. 

\subsection{Convolutional Neural Network}
\begin{figure*}[t]
{\centering
\includegraphics[width=12cm]{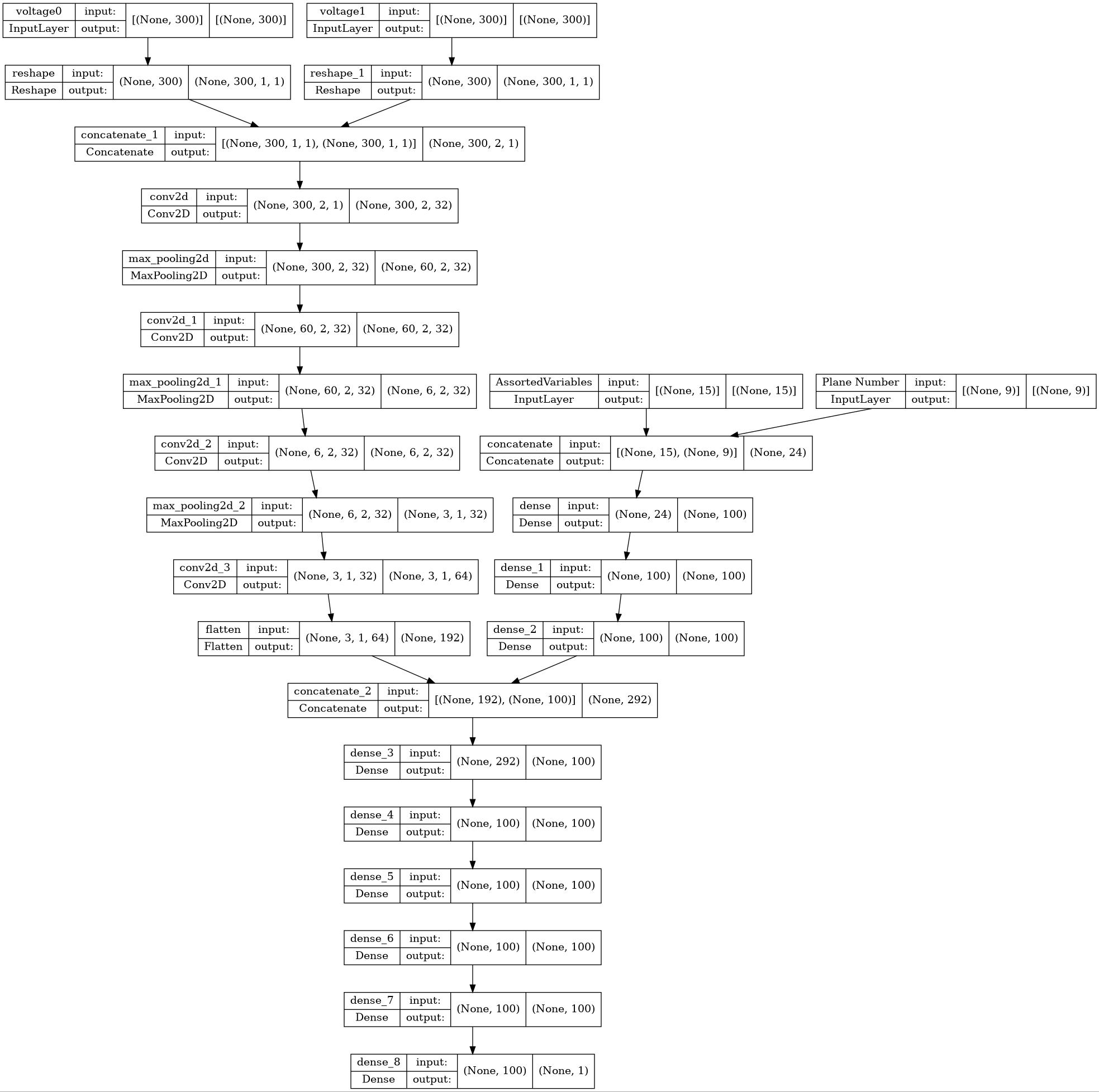}
\caption{The convolutional neural network architecture. It estimates the time-distance relationship using the hit time and the CDCH digitized voltages. 
\label{CNN}
}}\end{figure*}

All available information about the time distribution of the clusters in a hit is stored in the digital waveform. Here, the neural network is trained to "learn" how to extract the useful information from the waveform in order to optimize the DOCA estimate. The convolutional layers take advantage of the temporal relationship between waveform bins. We use 300 bins of data per waveform (300/0.6 GHz $\sim$ 500 ns). The starting bin is 60 ns before the estimated track time ($T_{0}$) at the hit. The model is initiated by reshaping the binned waveform data into a 300x2 array such that, for each of the 300 time bins, there are two waveform values (upstream/downstream wire ends). The convolutional neural network is shown in Figure \ref{CNN}. 

\section{Results}
\label{Results}
In this section, we compare results for the conventional, DNN, and CNN DOCA estimators. 
All results in this section use independent test data to avoid any training sample dependence. For all three DOCA estimators, the DOCAs are re-evaluated after the first pass of tracking and the tracks are refit. The first pass analysis uses the conventional analysis. We refit the tracks in all three cases for consistency. 

\begin{figure*}[t]
{\centering
\includegraphics[width=12cm]{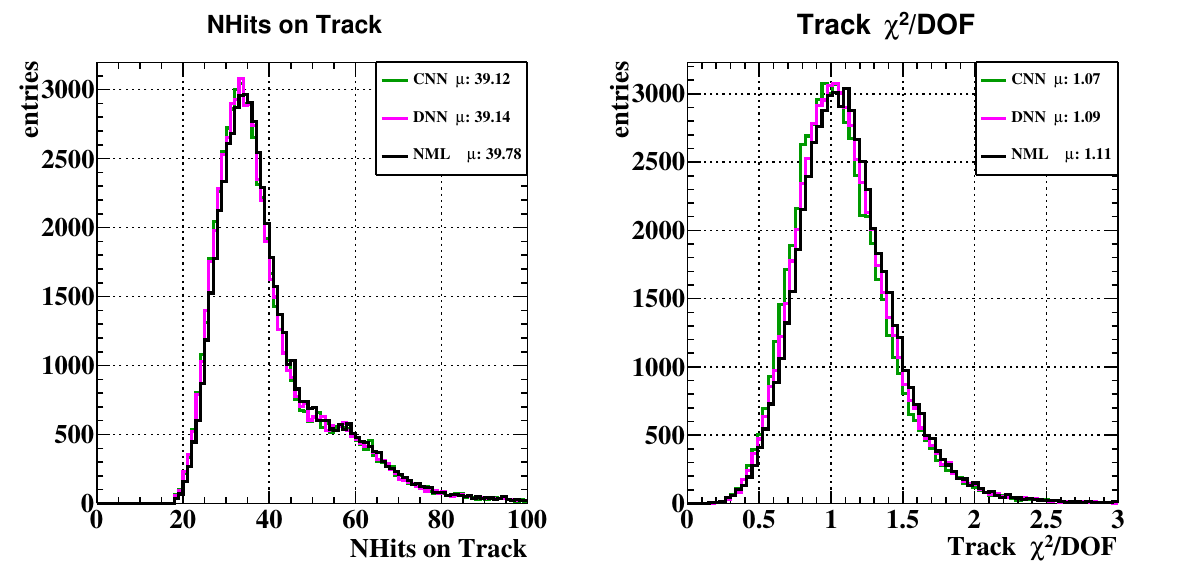}
\caption{The number of hits per track and the $\chi^{2}/DOF$ are compared in the three techniques. The conventional DOCA estimate uses a larger $\sigma_{DOCA}$ in the track fitting in order to get a comparable $\chi^{2}/DOF$ and number of hits. NML represents the standard or conventional DOCA estimation, DNN represents the dense neural network, and CNN represents the convolutional neural network that uses the full waveform information . 
\label{chi2}
}}\end{figure*}

The Kalman filter requires both the DOCAs and uncertainty ($\sigma_{DOCA}$)  to fit the track. The $\sigma_{DOCA}$ is used by the Kalman filter both for the track fit and for removing outliers. For all time-distance approaches, we use a 4th order polynomial to fit the squared DOCA residuals as a function of track DOCA to estimate $\sigma_{DOCA}$.  We then scale the $\sigma_{DOCA}$ constant parameter in the fit to achieve a comparable number of hits per track and a comparable $\chi^{2}/DOF$ (Figure \ref{chi2}). We use the same $\sigma_{DOCA}$ for both neural networks. This yields nearly the same tracking efficiency with a fixed track selection; facilitating the comparison among the different approaches. The neural network $\sigma_{DOCA}$ is $\sim$ 9\% smaller than that of the conventional approach. Alternatively, using the same $\sigma_{DOCA}$ in the three cases results in a lower tracking efficiency, less hits per track and a larger $\chi^{2}$ in the conventional approach. The results using the same $\sigma_{DOCA}$ are compared in Appendix \ref{SameSigDOCA}.


The description of results of the neural network analysis applied to MEG II data are organized as follows. Section \ref{HitResults} compares the hit residuals for the three techniques. Section \ref{TXY functionResults} compares the Garfield++ TXY function to that of the DNN. In Section \ref{DTResults}, a  data-driven positron measurement resolution technique is used to compare the DOCA estimator's effect at the kinematic level. The full MEG II analysis is ongoing and improvements in hit resolutions and kinematic measurements presented here may change as improvements to the MEG II analysis (e.g. tracking, wire alignment, magnetic field mapping, etc.) are incorporated. The 2021 MEG II detector performances are given in \cite{MEGIIDetector}.  

\subsection{Hit-Level Results}
\label{HitResults}

\begin{figure*}[t]
{\centering
\includegraphics[width=12cm]{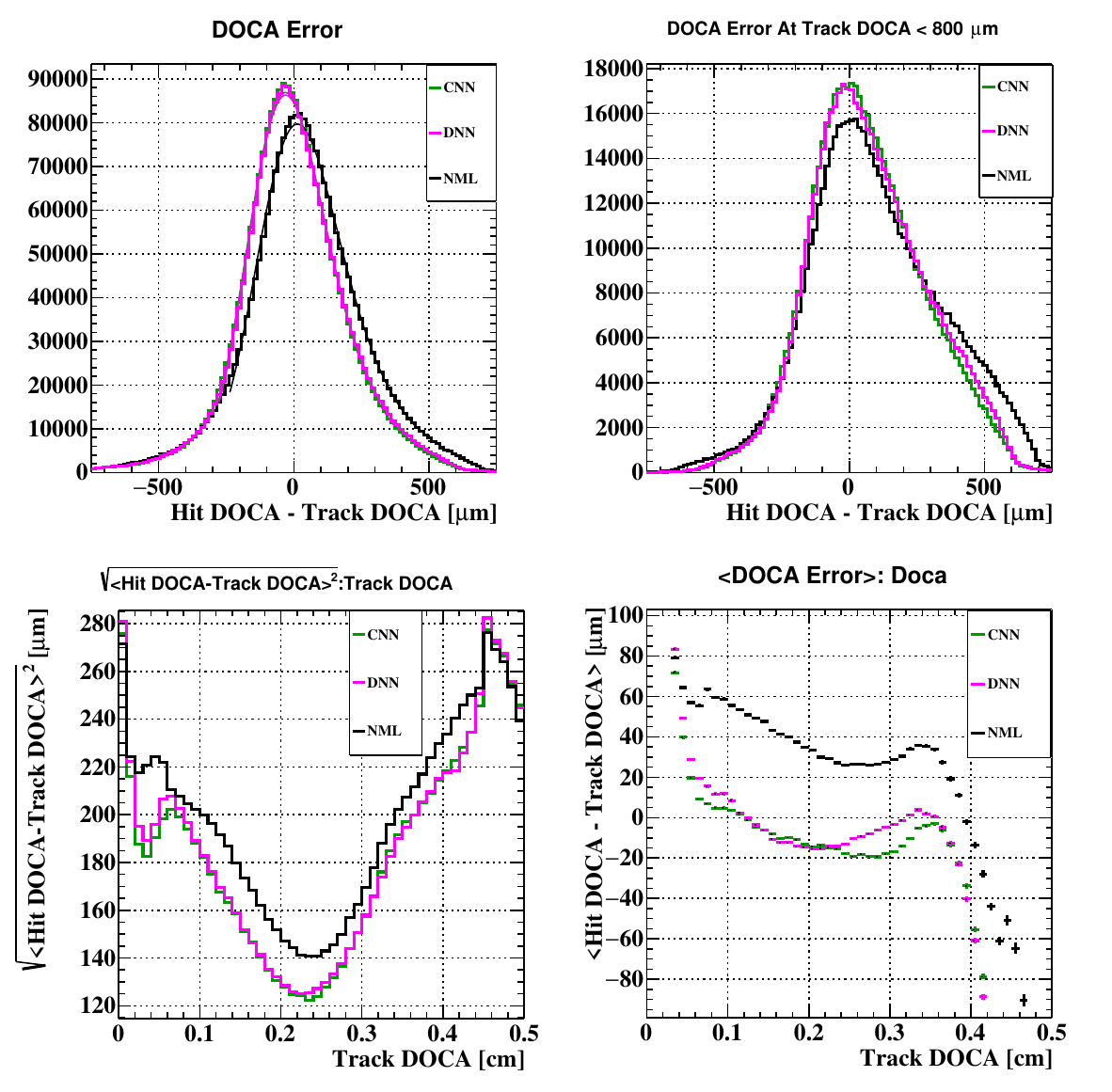}
\caption{The hit level results for the three DOCA estimators. NML represents the standard or conventional DOCA 
estimation, DNN represents the dense neural network, and CNN represents the convolutional neural network that includes the full waveform information. 
\label{Hit}
}}\end{figure*}


In this section, we present and discuss the results at the hit residual level, shown in Figure \ref{Hit}. The DOCA residual distributions for all approaches are shown top-left. The histograms are well-fit to double Gaussian functions with a core sigma of $114 \mu$m ($119 \mu $m) and a tail sigma of
  $236 \mu$m ($259 \mu$m) with a core fraction of 0.59 (0.70) for the convolutional neural network (conventional) approaches.
As expected, all distributions have a positive tail due to the asymmetry created by the ionization statistics (track is closer to the wire than the first ionization site) and instances when the first cluster is not detected in the waveform analysis.
Both neural network techniques result in less bias, with a slightly negative mode. The CNN has a slight reduction in the positive tail and a slight increase in the number of entries in the core compared to those for the DNN. We infer that the CNN uses the full waveform information to further suppress the ionization statistics bias, whereas the DNN is simply removing the average bias. In training, the CNN resulted in a $\sim3\%$ smaller loss with respect to the DNN ($\sum_{i}{|Y_{i}-\hat{Y_{i}}|}$). 
In the top-right, we show the DOCA residual distribution near the wire (track DOCA $< 800 \mu m$). Here there is the largest ionization statistics bias and also the largest improvement seen by the neural network approaches.   


We show the root mean squared residuals as a function of DOCA in the bottom-left. To avoid the effect of outliers, hits with a residual larger than $500 \mu m$ were excluded from the calculation. The improvement due to the two networks is largest($\sim18\%$) close to the wire
due to the larger ionization statistics bias there.  The CNN does better than the DNN mostly close to the wire (track DOCA < 1 mm). 

We show the <hit DOCA - track DOCA> (DOCA bias) as a function of track DOCA for the three techniques in the bottom-right. The non-zero mean DOCA bias in the conventional case is comparable to the DOCA bias expected from the MC from ionization statistics (Figure \ref{DocaBiasMC}); this is a good indication that, at least on average, we are correctly estimating the TXY function. In the CNN and DNN cases the <DOCA bias> has been highly suppressed in the middle of the cell, with a bias between $\sim[-20,20] \mu m$ for all hits with track DOCA between $\sim [0.06,0.38]cm$. 

Since both the hit and track DOCA are positive definite, when the track DOCA is close to zero, the only possibility is that the residual (hit DOCA - track DOCA) is positive. Hence, we expect a large bias simply due to this effect. 
The opposite is true at the edge of the cell. The neural network DOCA estimates should always return a value inside the cell to minimize the loss function and therefore we expect a negative <DOCA bias>. 

In Appendix \ref{BiasSuppression}, we discuss whether a simpler technique can suppress the DOCA bias in the conventional TXY function approach by using the transformation: $t_{drift} \rightarrow t_{drift} - 2ns$. On average, this corrects the bias and gives a <DOCA bias> similar to that of the the dense neural network. However, there are still fewer hits on tracks at small DOCA and a positive tail in the hit DOCA - track DOCA distribution with respect to the NN approaches.

\begin{figure*}[b]
{\centering
\includegraphics[width=16cm]{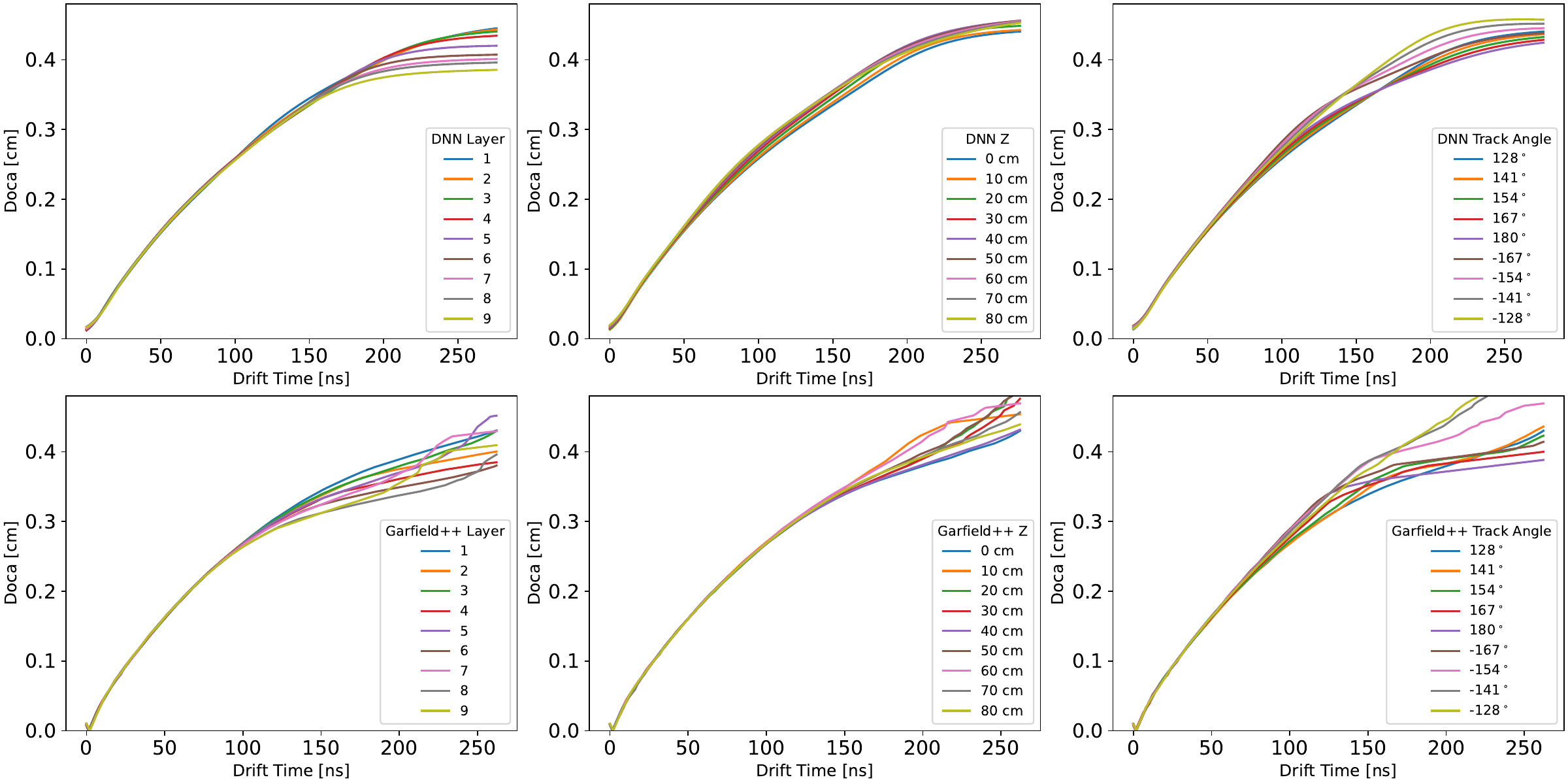}
\caption{The time-distance relationship as a function of layer (column 1), position along the wire axis, (column 2), and track angle (column 3) are shown for the dense neural network (top row, DNN) and the conventional Garfield++ TXY function (bottom row, No ML). 
\label{TXY functionFunction}
}}\end{figure*}

\subsection{Time-Distance Comparison}
\label{TXY functionResults}
In this section, we compare the TXY function from the conventional Garfield++ calculation\cite{CDCHDetector} to that of the dense neural network as a function of the layer number, the position along the wire axis (z), and the track angle. This comparison is shown in Figure \ref{TXY functionFunction}. Since the convolutional network requires digital waveforms to make a DOCA estimate, it's difficult to compare the TXY function in the same way. 
As a reminder, both layer number and position along the wire axis affect the cell size.

The Garfield++ TXY function results in non-smooth functions near the edges of the cell. This may be an effect of the positions of the cathode wires. The TXY function dependence on the CDCH layer is suppressed in the DNN case. 

Comparing the DNN case to that of Garfield++, the hit's position along the wire has a larger effect on the TXY function at drift times around $\sim 100$ ns, but a smaller effect at the edges of the cell. This implies that the dense neural network "learned" a different TXY function dependence on cell size than that simulated by Garfield++. 

Finally, we observe that the TXY function dependence on track angle is significantly different at large drift times ($> 150$ ns). This is likely the result of smoothing between the anode cathode geometry over many cells. An interesting point is that both the Garfield++ TXY and the DNN TXY show a significant difference between positive and negative track angles once reaching a drift time of $\sim 120$ ns. 

We conclude that the observed improvement in the root mean squared residuals and a suppressed DOCA bias at all track DOCA (Figure \ref{Hit}) indicates that the dense network produces the more accurate TXY function. 

\subsection{Evaluating the Effect on Positron Kinematic Measurements}
\label{DTResults}

In this section we verify that the improved DOCA resolution improves the resulting $e^{+}$ measurements. 

\begin{figure*}[b]
{\centering
\includegraphics[width=14cm]{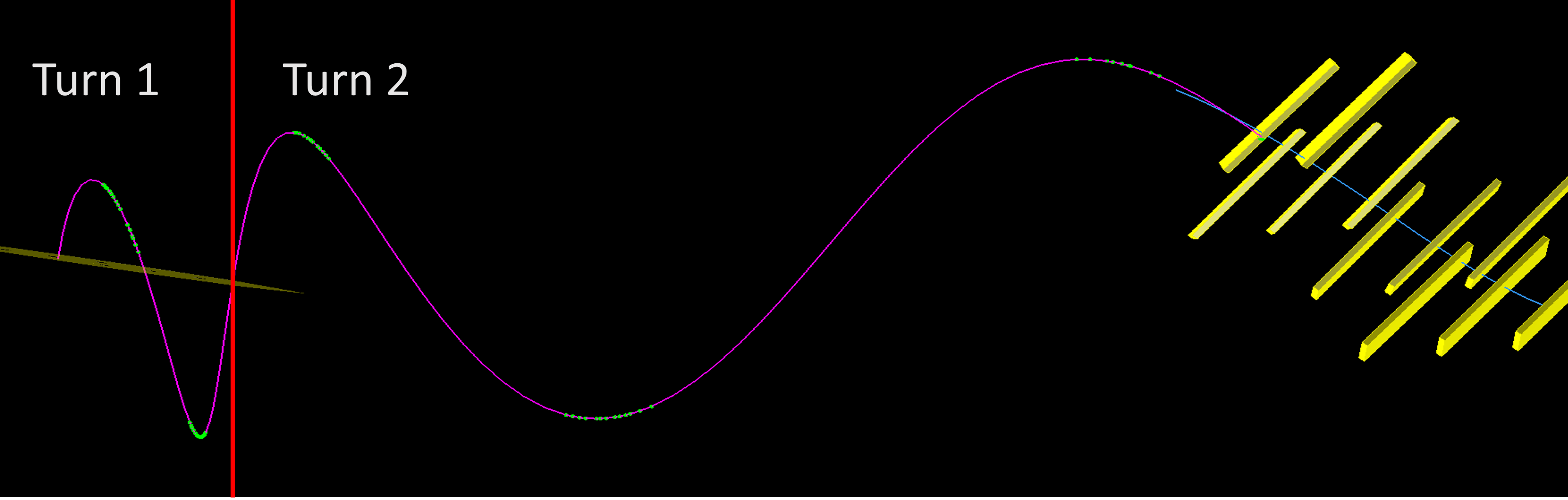}
\caption{An example of a double turn positron track. The green dots show intersected wires with signal in the drift chamber; the yellow tiles show the pixelated timing counter tiles with signal.
\label{DT}
}}\end{figure*}

In the MEG II experiment, $\sim 15\%$ of $e^{+}$ tracks pass through the drift chamber 5 times (9x5 layers intersected). An example double turn event is shown in Figure \ref{DT}. For these tracks, the first turn (2x9 layers intersected) and
the second turn (3x9 layers intersected) are used to independently fit and measure two state vectors. These are propagated, one forward and one backward, to the extended target plane between the two turns. The comparison of the two state vectors on a common plane is used to measure the CDCH performances, which is cross checked by the full detector simulation. This double turn analysis method was originally developed and implemented in the MEG I experiment.\cite{MEGI}. 

We use the double turn analysis method to estimate the relative $e^{+}$ track measurement quality using the three DOCA estimators. Improved agreement between the kinematics implies improved kinematic precision. We fit each histogram to the convolution of two double Gaussians. We fix the two double Gaussians to be identical.

These distributions are not the MEG II signal positron resolutions, but the resolution of the first and second turns of Michel positron tracks added in quadrature. The first turn only intersects the chamber two times and thus has a degraded resolution with respect to the standard Michel track (three intersections). Extracting out the eventual MEG II signal resolutions requires corrections from the Monte Carlo simulation. Signal positron resolutions extracted using the double turn analysis are described here\cite{CDCHDetector}. Nonetheless, this fit gives kinematic precision estimates that can be used to compare the DOCA estimators.  


\begin{figure*}[b]
{\centering
\includegraphics[width=14cm]{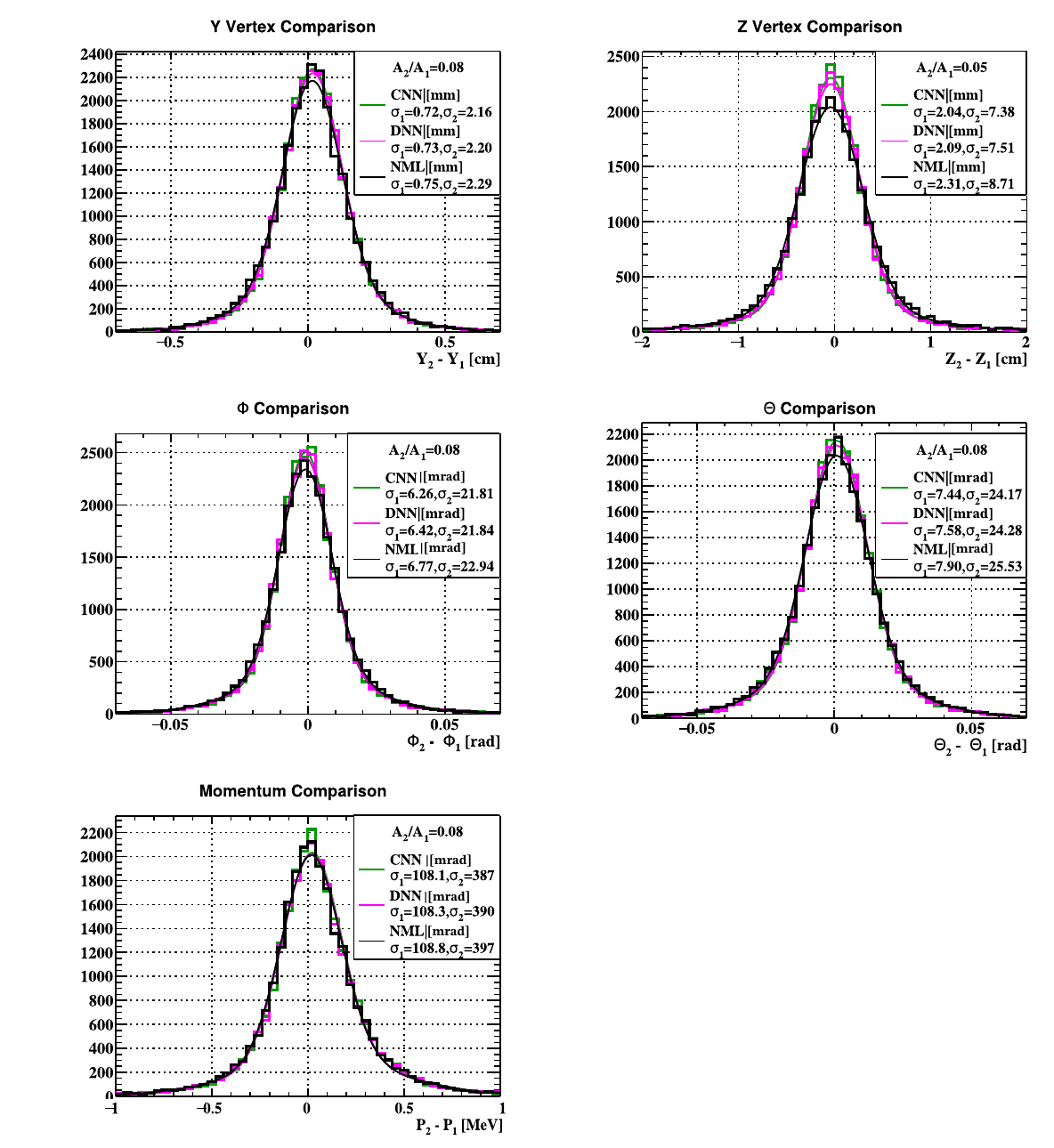}
\caption{ Here, we show the kinematic comparison at a common plane between two independently fit and measured turns of a single two-turn track. NML represents the standard or conventional DOCA estimation, DNN represents the dense neural network, and CNN represents the convolutional neural network that inputs waveform voltages. These distributions are preliminary and do not represent the signal positron resolution, but the resolution of the first and second turns of Michel positron tracks added in quadrature.
\label{DoubleTurn}
}}\end{figure*}

All positron resolutions have improved when incorporating either neural network DOCA estimator. Here, the relative amplitude of the core and the tail of the two Gaussians are fixed for all DOCA estimators. When comparing the conventional approach to the convolutional neural network, the CNN core resolutions improve by  $4.0\%$, $11.7\%$, $7.5\%$, $5.8\%$, $0.6\%$ for $y_{e}$, $z_{e}$, $\phi_{e}$, $\theta_{e}$,  $p_{e}$ respectively. The uncertainties in the core $\sigma$ are 0.0073 mm, 0.019 mm, 0.066 mrad, 0.076 mrad, and 1.16 keV for $y_{e}$, $z_{e}$, $\phi_{e}$, $\theta_{e}$,  $p_{e}$ respectively, well below the differences between the network resolutions and the conventional resolutions. In MC studies, improved DOCA resolutions had the largest fractional improvement on the $z_{e}$ kinematic resolution; this supports our results, which also show the largest fractional improvement in $z_{e}$. The tails are also wider in all kinematic distributions in the conventional approach. In Appendix \ref{SameSigDOCA}, we note that if the same $\sigma_{DOCA}$ is used in all cases, the double turn analysis results in less entries in the peak and degraded resolutions with respect to the neural networks. We also mention that the double turn resolutions improve when using the CNN instead of the DNN. This implies that the network's use of the waveform results in an improvement at the kinematic level, though this is a minor effect. 

\section{Potential Improvements}
\label{Discuss}
In this section, we discuss potential improvements to the networks and their use in particle tracking software, some particular to MEG II and some more general. 
In the MEG II application, the algorithm has access to the unique layer number through a hot encoded array (Figure \ref{CNN}). Alternatively, the neural network could input a hot encoded array containing the unique wire number. This would require significantly more training data (factor of $\sim 100$) in order to have the same precision i.e. creating the TXY function for each wire as a function of drift time, position along the wire, track angle, etc. This would allow the network to learn wire-specific subtleties such as the relative anode cathode geometry or signal/noise. 

In some applications, a hybrid application using both the conventional and the neural network TXY functions might improve performance. This was tested for the MEG II case by including the Garfield++ based DOCA estimate as an additional neural network input variable. This did not yield a significant performance improvement, but might lead to improvements for other experiments.  

In the MEG II application we found that a single convolutional neural network could achieve a DOCA estimate without using the waveform analysis-driven hit arrival time but performance was worse than even that of the conventional approach. 
Achieving optimal resolution without the waveform analysis might be possible, but developing such an analysis is beyond the scope of this paper. It may require a multi-step neural network. For example, a first step that finds the hit arrival time and then a second step of converting that into a DOCA estimate. This was tried in MEG II and is discussed here\cite{CDCHDetector}.

Another possible approach to improve performance is to use a hybrid of conventional cluster counting (e.g. finding peaks in the waveform) with machine learning. This could involve adding a list of the waveform peak times (possibly not resolving all clusters) as an additional input to the neural network. Just as the waveform analysis hit time is required to have the optimal neural network, maybe the found peak times could improve the network further. 

Finally, we return to the point that our training was optimized by using low intensity beam rate data. The improvement using low intensity data implies that the neural networks are not able to learn how to distinguish between the contributions from in-time and out of time (pileup) tracks to the waveform in the training process.
One possible method to improve the performance is to implement a conventional pileup detection algorithm that produces a binary flag indicating evidence of more than one hit on the waveform. By developing separate machine learning algorithms for the two cases, the network might improve performance for the signal DOCA in the pileup case. 


\section{Conclusions}
In this paper, we describe the application of neural network models to improve the distance of closest approach measurement in drift chambers. The application results in a data-driven time-distance relationship, unlike conventional approaches. The data-driven approach accounts for all detector-specific properties (e.g. longitudinal diffusion, gas properties, ionization statistics, electronic noise, etc.). 
 In an example application to the MEG II drift chamber data, the networks improve the track position measurements with respect to those of the conventional approach. The optimal configuration is a convolutional neural network that uses the signal waveforms. This gives the network access to information from all ionization clusters, which is not conventionally used in DOCA estimation.

\section{Acknowledgements}    
We thank the members of the MEG II collaboration. In particular, we thank the CDCH hardware group for developing and operating the drift chamber through the 2021+2022 physics runs. We thank Marco Chiappini for his contributions to the determination of the alignment parameters of the CDCH. We thank Yusuke Uchiyama for help with the CDCH waveform analysis, machine learning implementation, and discussions. We thank Fedor Ignatov for his work on the track finding and track fitting software. We thank Francesco Renga and Franco Grancagnolo  for useful discussions in the MEG II positron analysis group. We thank Atsushi Oya for his contributions to the double turn analysis of the MEG II data.  We are grateful for the support and co-operation provided by PSI as the host laboratory. This work was supported by U.S. Department of Energy grant DEFG02-91ER40679.

\appendix
\section*{Appendices}
\addcontentsline{toc}{section}{Appendices}
\renewcommand{\thesubsection}{\Alph{subsection}}
\setcounter{table}{0}
\renewcommand{\thetable}{A\arabic{table}}
\setcounter{figure}{0}
\renewcommand{\thefigure}{A\arabic{figure}}

\section{Positron Measurements Using The Same $\sigma_{DOCA}$} 
\label{SameSigDOCA}

As a check, we verify that the double turn measurements are still improved in the neural network cases when using the same Kalman filter $\sigma_{DOCA}$ with all DOCA estimators. Here, the resolution when using the conventional TXY function has effectively stayed the same, and in some kinematics (e.g. momentum comparison) the resolution is degraded. Here, we adjusted the track selection on the number of hits per track to achieve a comparable tracking efficiency. 
\begin{figure*}[ht]
{\centering
\includegraphics[width=14cm]{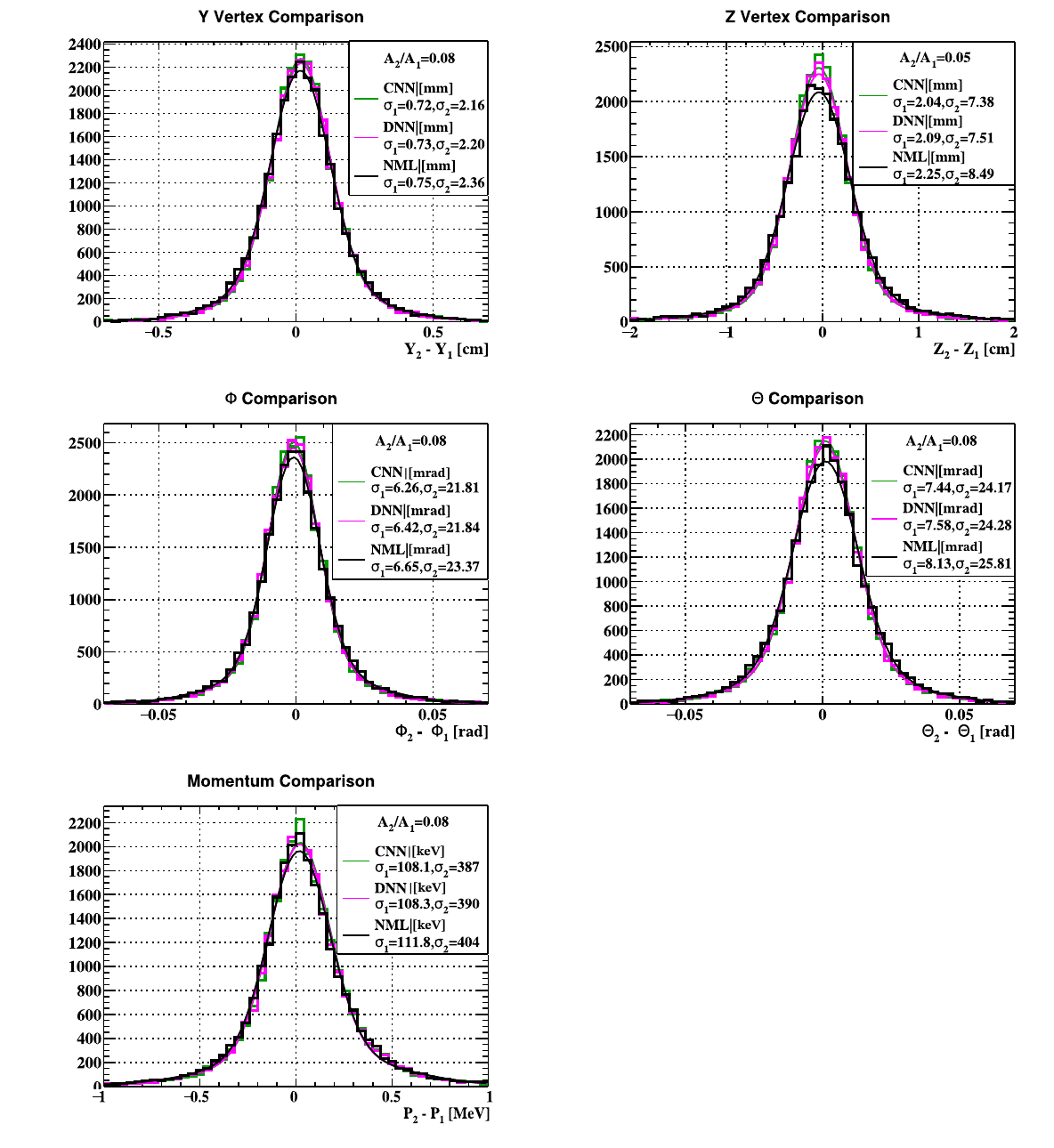}
\caption{Here, we show the kinematic comparison at a common plane between two independently fit/measured turns of a single two-turn track. In this plot, all fits use the same Kalman filter $\sigma_{DOCA}$. The NML case represents the standard or conventional DOCA estimation, DNN represents the dense neural network, and CNN represents the convolutional neural network that inputs waveform voltages. 
\label{DTSameSig}
}}\end{figure*}

\section{Positron Kinematic Measurements When Suppressing the Bias in the Conventional TXY function} 
\label{BiasSuppression}
\renewcommand{\thesubsection}{\Alph{subsection}}
\setcounter{table}{0}
\renewcommand{\thetable}{B\arabic{table}}
\setcounter{figure}{0}
\renewcommand{\thefigure}{B\arabic{figure}}

\begin{figure*}[ht]
{\centering
\includegraphics[width=12cm]{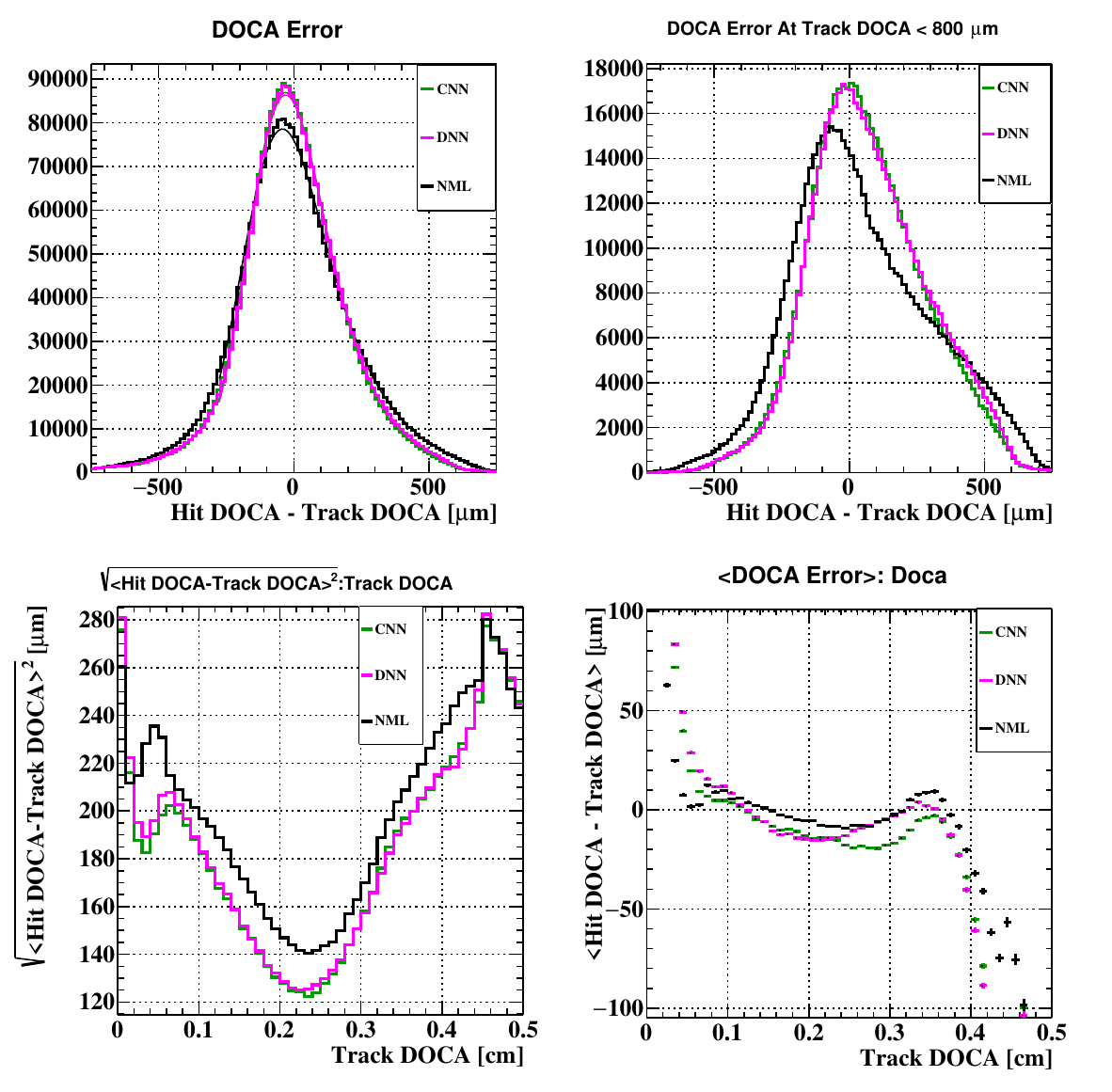}
\caption{The hit level results are compared using the three DOCA estimators. In this plot, the conventional TXY function has been modified to suppress the DOCA bias; transforming $t_{drift} \rightarrow t_{drift} - 2ns$. NML represents the standard or conventional DOCA estimation, DNN represents the dense neural network, and CNN represents the convolutional neural network that includes the waveform voltages. 
\label{DocaBias-2}
}}\end{figure*}

The conventional analysis results in a <DOCA bias> of $\sim 50 \mu m$; a possible way to correct for this is a simple correction to the drift time to remove this bias.
The conventional TXY function was modified, transforming $t_{drift} \rightarrow t_{drift} - 2ns$. In Figure \ref{DocaBias-2}, the hit resolutions are compared again in the three cases. 

\begin{figure*}[ht]
{\centering
\includegraphics[width=14cm]{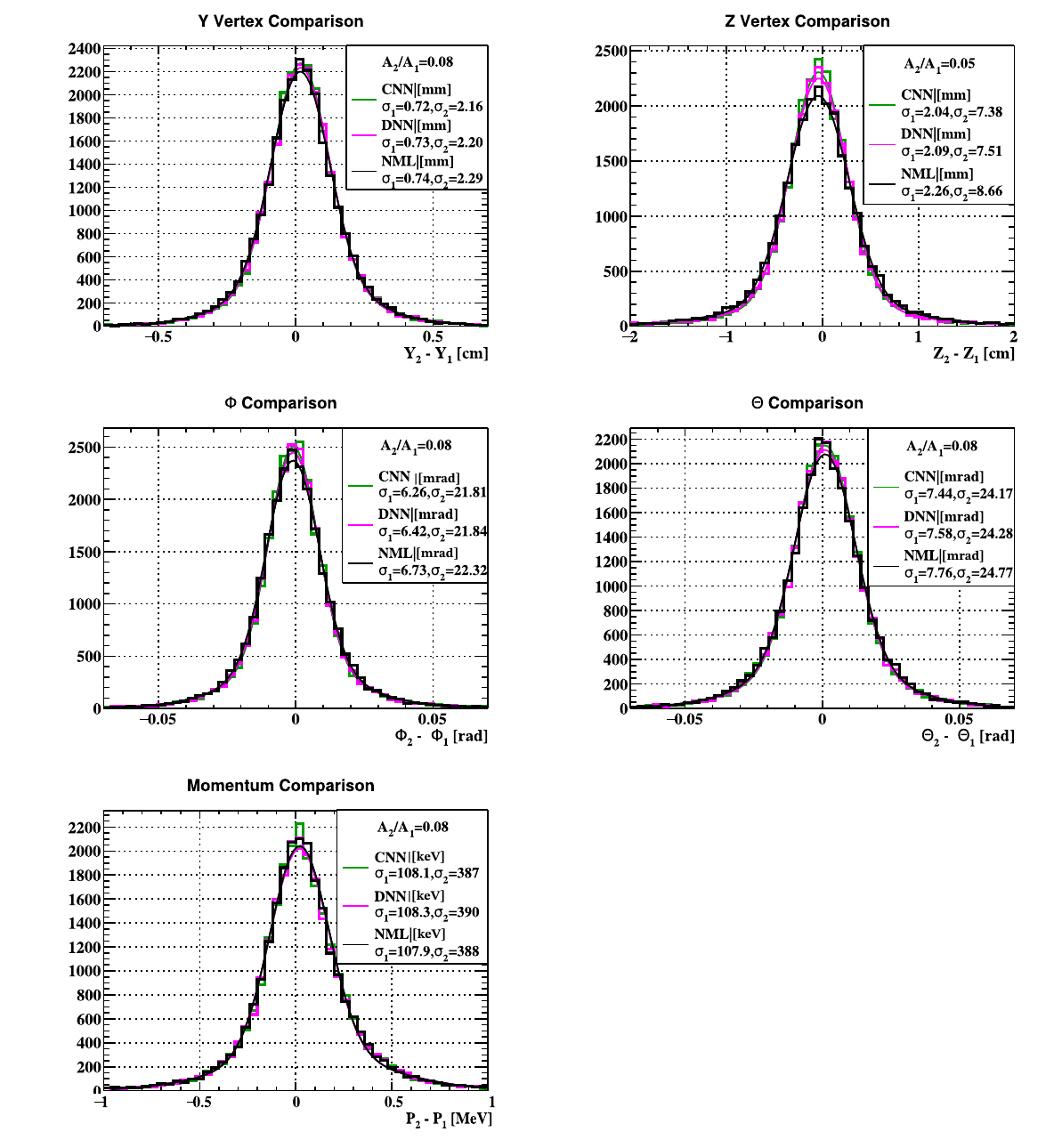}
\caption{Here, we show the kinematic comparison at a common plane between two independently fit/measured turns of a single two-turn track. In this plot, the conventional TXY function has been modified to suppress the DOCA bias; transforming $t_{drift} \rightarrow t_{drift} - 2ns$. The NML case represents the standard or conventional DOCA estimation, DNN represents the dense neural network, and CNN represents the convolutional neural network that inputs waveform voltages. 
\label{DTDocaBias-2}
}}\end{figure*}

In Figure \ref{DTDocaBias-2}, we compare the positron double turn measurement resolutions with the modified conventional TXY function. We note that just suppressing this <DOCA bias> in the conventional approach does indeed yield a slight resolution improvement; the convolutional neural network case still results in the optimal resolution. 

Finally, using the same $\sigma_{DOCA}$ as the neural network approach and applying the $t_{drift} \rightarrow t_{drift} - 2ns$ transformation also results in comparable measurements.

\end{document}